\documentclass[prl,showpacs,preprint]{revtex4}%
\usepackage{amsmath}
\usepackage{graphicx}
\usepackage{amsfonts}
\usepackage{amssymb}%
\begin{document}
\title{Novel interferometer to beat the standard quantum limit using optical
transverse modes in multimode waveguide}
\author{Jian Fu}
\affiliation{State Key Lab of Modern Optical Instrumentation, Department of Optical
Engineering, Zhejiang University, Hangzhou 310027, China}

\begin{abstract}
We propose a novel interferometer by using optical transverse modes in
multimode waveguide that can beat the standard quantum limit. In the scheme,
the classical simulation of $N$-particle quantum entangled states is generated
by using $N$ independent classical fields and linear optical elements. Similar
to the quantum-enhanced measurements, the classical simulation can also
achieve $\sqrt{N}$ enhancement over the precision of the measurement $N$ times
for independent fields. Due to only using classical fields and linear optical
elements, the scheme can be realized much more easily.

\end{abstract}

\maketitle

Optical interferometric techniques are in particular widely used in ultimate
sensitivity measurements, such as gravity-wave detection \cite{Thorne},
nanometric displacement measurement \cite{Kamimura} and optical gyroscopes
\cite{Chou}. In these measurements, the ultimate sensitivity is conventionally
bounded by the quantum nature of the electromagnetic field. It has been shown
that the so-called shot noise or standard quantum limit are due to the vacuum
fluctuations coupled to the interferometer and to the random motion of the
mirrors induced by the radiation pressure fluctuations \cite{Caves}. However,
these conventional limits are not as fundamental as the Heisenberg limits
\cite{Holland}, and can be beaten by using quantum entanglement
\cite{Giovannetti} and squeezing \cite{Xiao}. Quantum entanglement has been
proved to allow a precision enhancement equal to the square root of the number
$N$ of employed particles, which can achieve the Heisenberg limits
\cite{Giovannetti}. But there is an enormous difficulty in the
quantum-enhanced measurement, which is usually very complicated to realize
multi-particle quantum entanglement even as few as 5 or 6 particles
\cite{Zhao}.

Recently, \textquotedblleft mode-entangled states\textquotedblright\ based on
the transverse modes of classical optical fields propagating in multimode
waveguides are proposed as classical simulation of quantum entangled states
\cite{Fu}. It is interesting that the mode-entangled states can also exhibit
the nonlocal correlations, such as the violation of Bell's inequality. The
states can be regarded as the nonlocal generalization of the transient
interference effect between two independent laser beams \cite{Magyar} and
explained by a random phase ensemble model based on classical electromagnetics
\cite{Fu1}. The simulation not only helps to understand the nonlocal
properties of quantum entanglement from a new viewpoint, but also arouses
interest in a full optical quantum computation scheme based on the transverse
modes of classical fields \cite{Fu2,Man'ko}.

In this letter, we will propose a novel interferometer to beat the standard
quantum limit using mode-entangled states. Similar to the quantum-enhanced
measurements, the $N$-field mode-entangled states can also achieve $\sqrt{N}$
enhancement over the precision of the measurement $N$ times for independent
fields. Moreover, the interferometer can be realized more easily than the
quantum-enhanced measurement due to only using classical fields and linear
optical elements. Before going into the scheme of the new interferometer, we
would like to introduce an ordinary interferometer using optical transverse
modes in multimode waveguide.

Considering a weakly guiding, symmetric slab waveguide, an optical field in
the propagation $z$ direction is restricted within the core region, which has
the higher refractive index (RI) compared with that of the cladding. We assume
that a dual-mode waveguide support two normal modes, namely
$\operatorname*{TE}\nolimits_{0}$ mode and $\operatorname*{TE}\nolimits_{1}$
mode. Thus the coherent superposition state of the two modes can be described
as
\begin{equation}
\left\vert \psi\right\rangle =\binom{C_{0}e^{i\beta_{0}z}}{C_{1}e^{i\beta
_{1}z}}, \label{e1}%
\end{equation}
where $\beta_{0}$ and $\beta_{1}$ are the propagation constants of the modes
$\operatorname*{TE}\nolimits_{0}$ and $\operatorname*{TE}\nolimits_{1}$,
respectively. Apparently, a mode analyzer (MA) that contains a variable phase
modulator $\theta$ and Y splitter proposed in \cite{Fu} is a simple dual-mode
waveguide interferometer. We can define an intensity difference operator for
the MA's outputs,
\begin{equation}
\hat{A}\left(  \theta\right)  =\left(
\begin{array}
[c]{cc}%
0 & e^{i\theta}\\
e^{-i\theta} & 0
\end{array}
\right)  . \label{e2}%
\end{equation}
When the input field of MA is prepared in the mode superposition $\left\vert
\psi\right\rangle $, the intensity difference of the MA's outputs can be
obtained
\begin{align}
A\left(  \theta\right)   &  =\left\langle \psi\right\vert \hat{A}\left(
\theta\right)  \left\vert \psi\right\rangle \label{e3}\\
&  =C_{0}C_{1}^{\ast}e^{i\left(  \Delta\beta z+\theta\right)  }+C_{1}%
C_{0}^{\ast}e^{-i\left(  \Delta\beta z+\theta\right)  },\nonumber
\end{align}
with $\Delta\beta=\beta_{1}-\beta_{0}$. When $C_{0}=C_{1}=1/\sqrt{2},A\left(
\theta\right)  =\cos\left(  \Delta\beta z+\theta\right)  .$ By using error
propagation theory and Eq. (\ref{e3}), it is very easy to evaluate an overall
phase error $\Delta\theta=1/\sqrt{N}$ for repeating the experiment $N$ times,
which is the standard quantum limit for the interferometers.

In the quantum-enhanced measurements, $N$-particle quantum entangled states
are required to achieve $\sqrt{N}$ precision enhancement. Similarly, in our
scheme, $N$-field mode-entangled states are required. By using numerical
simulation, the CNOT gate scheme for generating mode-entangled states has
proved feasible in Ref. \cite{Fu1}. Here we propose a new scheme using only
linear optical elements to realize the classical simulation of quantum
entanglement. In the scheme, by using a properly designed directional coupler
(DC), two independent classical fields prepared in mode superpositions are
completely exchanged $\operatorname*{TE}\nolimits_{0}$ mode or
$\operatorname*{TE}\nolimits_{1}$ mode. The two output fields of the DC can
exhibit the violation of Bell's inequality in the correlation measurement
scheme proposed in Ref. \cite{Fu}.

If classical fields are quasi-monochromatic (i.e., the spread $\Delta k$ is
much less than the midwave number $k_{0}$, $\Delta k\ll k_{0}$), two fields
are statistically independent, as mainly reflected in their independent random
phases $\phi_{a},\phi_{b}$ uniformly distributed in $\left[  0,2\pi\right]  $.
Then the states of two independent fields prepared in the mode superpositions
are defined as
\begin{equation}
\left\vert \psi_{a}\right\rangle =e^{i\phi_{a}}\binom{C_{0}^{a}}{C_{1}^{a}%
},\left\vert \psi_{b}\right\rangle =e^{i\phi_{b}}\binom{C_{0}^{b}}{C_{1}^{b}},
\label{e4}%
\end{equation}
where $C_{0,1}^{a},C_{0,1}^{b}$ are the mode coefficients of two optical
fields, respectively. By properly adjusting the coupling coefficient and
length, we can design a dual-mode waveguide DC to realize mode separating or
combining \cite{Fu2}. By using the finite differential beam propagating method
(FD-BPM) \cite{Yevick}, we have simulated numerically mode separating by using
the properly designed DC. The result is shown in Fig. 1. When the two fields
$\left\vert \psi_{a}\right\rangle ,\left\vert \psi_{b}\right\rangle $ are
respectively sent into two inputs of the DC to completely exchange their
$\operatorname*{TE}\nolimits_{1}$ modes, we obtain the output two fields that
have become incoherent mode superpositions,
\begin{equation}
\left\vert \psi_{a}^{\prime}\right\rangle =e^{i\phi_{a}}\binom{C_{0}^{a}%
}{C_{1}^{b}e^{i\lambda}},\left\vert \psi_{b}^{\prime}\right\rangle
=e^{i\phi_{b}}\binom{C_{0}^{b}}{C_{1}^{a}e^{-i\lambda}}, \label{e5}%
\end{equation}
where $\lambda=\phi_{b}-\phi_{a}$ is the random phase difference uniformly
distributed in $\left[  0,2\pi\right]  $. The output state of the DC can be
written as the product of the two incoherent mode superpositions,
\begin{equation}
\left\vert \psi^{\prime}\right\rangle =\left\vert \psi_{a}^{\prime
}\right\rangle \otimes\left\vert \psi_{b}^{\prime}\right\rangle =e^{i\left(
\phi_{a}+\phi_{b}\right)  }\left(
\begin{array}
[c]{c}%
C_{0}^{a}C_{0}^{b}\\
e^{-i\lambda}C_{0}^{a}C_{1}^{a}\\
e^{i\lambda}C_{1}^{b}C_{0}^{b}\\
C_{1}^{a}C_{1}^{b}%
\end{array}
\right)  . \label{e6}%
\end{equation}
Assumed $C_{0,1}^{a},C_{0,1}^{b}=1/\sqrt{2}$, the density matrix $\rho$
describing the state is then obtained
\begin{equation}
\rho=\left\vert \psi^{\prime}\right\rangle \left\langle \psi^{\prime
}\right\vert =\frac{1}{4}\left(
\begin{array}
[c]{cccc}%
1 & e^{i\lambda} & e^{-i\lambda} & 1\\
e^{-i\lambda} & 1 & e^{-2i\lambda} & e^{-i\lambda}\\
e^{i\lambda} & e^{2i\lambda} & 1 & e^{i\lambda}\\
1 & e^{i\lambda} & e^{-i\lambda} & 1
\end{array}
\right)  . \label{e7}%
\end{equation}
Consider an ensemble of the independent and identical systems that are labeled
by the random phase $\lambda$ that satisfies normalization condition
$\int_{\Lambda}\Phi\left(  \lambda\right)  d\lambda=1$, where $\Phi\left(
\lambda\right)  $ is a distribution function and $\Lambda\subseteq\left[
0,2\pi\right]  $ is spanned by $\lambda$. Due to $\int_{\Lambda}e^{i\lambda
}\Phi\left(  \lambda\right)  d\lambda=0$, the density matrix $\rho$ can be
reduced by the ensemble average of $\lambda$,
\begin{equation}
\rho_{\lambda}=\frac{1}{4}\left(
\begin{array}
[c]{cccc}%
1 & 0 & 0 & 1\\
0 & 1 & 0 & 0\\
0 & 0 & 1 & 0\\
1 & 0 & 0 & 1
\end{array}
\right)  . \label{e8}%
\end{equation}
Apparently, the density matrix can not be factorized into the production of
separate density matrices and is also different from that of quantum entangled
state. If properly chose a operator $\hat{Q}$, we can obtain the same
expectation value of $\hat{Q}$ as quantum entangled states. In the correlation
measurement of Bell's inequality proposed in Ref. \cite{Fu}, the intensity
difference operators of MAs are this kind of operators. Therefore, we can
obtain the violation of Bell's inequality by using mode-entangled states. This
imply that the inseparability of the states similar to quantum entanglement
might be caused by a random phase mechanism. Here, by using the same
correlation measurement, we can obtain the normalized correlation function,
\begin{align}
S\left(  \theta_{1},\theta_{2}\right)   &  =\frac{\left\langle
\operatorname*{Tr}\left[  \rho\hat{A}\left(  \theta_{1}\right)  \hat{B}\left(
\theta_{2}\right)  \right]  \right\rangle _{\lambda}}{\sqrt{\left\langle
\operatorname*{Tr}\left[  \rho_{a}\hat{A}^{2}\left(  \theta_{1}\right)
\right]  \right\rangle _{\lambda}}\sqrt{\left\langle \operatorname*{Tr}\left[
\rho_{b}\hat{B}^{2}\left(  \theta_{2}\right)  \right]  \right\rangle
_{\lambda}}}\label{e9}\\
&  =\frac{\int_{\Lambda}\cos\left(  \theta_{1}+\lambda\right)  \cos\left(
\theta_{2}-\lambda\right)  \Phi\left(  \lambda\right)  d\lambda}{\sqrt
{\int_{\Lambda}\cos^{2}\left(  \theta_{1}+\lambda\right)  \Phi\left(
\lambda\right)  d\lambda}\sqrt{\int_{\Lambda}\cos^{2}\left(  \theta
_{2}-\lambda\right)  \Phi\left(  \lambda\right)  d\lambda}}\nonumber\\
&  =\cos\left(  \theta_{1}+\theta_{2}\right)  ,\nonumber
\end{align}
where $\left\langle ...\right\rangle _{\lambda}$ denote the ensemble averages
$\int_{\Lambda}...\Phi\left(  \lambda\right)  d\lambda$, and $\hat{A}\left(
\theta_{1}\right)  ,\hat{B}\left(  \theta_{2}\right)  $ are the intensity
difference operators of MAs operated on the fields $\left\vert \psi
_{a}^{\prime}\right\rangle $ and $\left\vert \psi_{b}^{\prime}\right\rangle $,
and the reduced density matrices $\rho_{a}$, $\rho_{b}$ are the partial traces
$\operatorname*{Tr}_{b}\left(  \rho\right)  $ and $\operatorname*{Tr}%
_{a}\left(  \rho\right)  $, respectively. Substituting the correlation
function into the Bell inequality \cite{Bell} (CHSH inequality \cite{CHSH}),
\begin{equation}
\left\vert B\right\vert =\left\vert S\left(  \theta_{1},\theta_{2}\right)
-S\left(  \theta_{1},\theta_{2}^{\prime}\right)  +S\left(  \theta_{1}^{\prime
},\theta_{2}^{\prime}\right)  +S\left(  \theta_{1}^{\prime},\theta_{2}\right)
\right\vert \leq2, \label{e10}%
\end{equation}
the violation can be obtained by proper choice of the phases $\theta_{1}$ and
$\theta_{2}$. By using FD-BPM and the method referred in Ref. \cite{Fu1}, we
numerically demonstrate the normalized correlation functions for the two
fields, as shown in Fig. 2. And the maximum violations of Bell's inequality
are obtained, as shown in Table 1, where the maximum values of $\left\vert
B\right\vert $ are the average results of many $\lambda$'s sequences.

Similarly, we can obtain the classical simulation of 3-particle GHZ state
\cite{GHZ} by using three independent classical fields and two DCs. First, two
fields are exchanged their $\operatorname*{TE}\nolimits_{1}$ modes by using
the first DC, then one of the output fields with the third field are exchanged
their $\operatorname*{TE}\nolimits_{1}$ modes by using the second DC. The
output fields can be obtained
\begin{equation}
\left\vert \psi_{a}^{\prime}\right\rangle =e^{i\phi_{a}}\binom{C_{0}^{a}%
}{C_{1}^{b}e^{i\lambda_{1}}},\left\vert \psi_{b}^{\prime}\right\rangle
=e^{i\phi_{b}}\binom{C_{0}^{b}}{C_{1}^{c}e^{i\lambda_{2}}},\left\vert \psi
_{c}^{\prime}\right\rangle =e^{i\phi_{c}}\binom{C_{0}^{c}}{C_{1}%
^{a}e^{i\lambda_{3}}}, \label{e11}%
\end{equation}
where $\lambda_{1}=\phi_{b}-\phi_{a},\lambda_{2}=\phi_{c}-\phi_{b},\lambda
_{3}=\phi_{a}-\phi_{c}$ are the phase differences, and $\phi_{a},\phi_{b}%
,\phi_{c}$ uniformly distributed in $\left[  0,2\pi\right]  $ are the random
phases and $C_{0,1}^{a},C_{0,1}^{b},C_{0,1}^{c}$ are the mode coefficients of
three optical fields, respectively. Then the three fields are sent to three
separated MAs denoted by $\hat{A}\left(  \theta_{1}\right)  ,\hat{B}\left(
\theta_{2}\right)  $ and $\hat{C}\left(  \theta_{3}\right)  $ respectively, we
obtain the correlation function for three intensity differences,
\begin{align}
S\left(  \theta_{1},\theta_{2},\theta_{3}\right)   &  =\left\langle \hat
{A}\left(  \theta_{1}\right)  \hat{B}\left(  \theta_{2}\right)  \hat{C}\left(
\theta_{3}\right)  \right\rangle \label{e12}\\
&  =%
{\displaystyle\iiint\nolimits_{\Lambda}}
A\left(  \theta_{1},\lambda_{1}\right)  B\left(  \theta_{2},\lambda
_{2}\right)  C\left(  \theta_{3},\lambda_{3}\right)  \Phi_{1}\left(
\lambda_{1}\right)  \Phi_{2}\left(  \lambda_{2}\right)  \Phi_{3}\left(
\lambda_{3}\right)  d\lambda_{1}d\lambda_{2}d\lambda_{3}\nonumber\\
&  =%
{\displaystyle\iiint\nolimits_{\Lambda}}
\cos\left(  \theta_{1}+\lambda_{1}\right)  \cos\left(  \theta_{2}+\lambda
_{2}\right)  \cos\left(  \theta_{3}+\lambda_{3}\right)  \Phi_{1}\left(
\lambda_{1}\right)  \Phi_{2}\left(  \lambda_{2}\right)  \Phi_{3}\left(
\lambda_{3}\right)  d\lambda_{1}d\lambda_{2}d\lambda_{3}\nonumber\\
&  =\frac{1}{4}\cos\left(  \theta_{1}+\theta_{2}+\theta_{3}\right)  ,\nonumber
\end{align}
where $\Phi_{i}\left(  \lambda_{i}\right)  $ are the distribution functions of
$\lambda_{i}$. Obviously, the correlation function is similar to that of GHZ
state except a normalization factor, that can also present Bell's theorem
without inequalities.

By using the linear optical scheme to generate mode-entangled states, we
propose a novel interferometer to beat the standard quantum limit, the scheme
is shown as Fig. 3. In the scheme, $N$ independent classical fields
$\left\vert \psi_{i}\right\rangle \left(  i=1...N\right)  $ are prepared in
mode superpositions. Then the $N$ fields are completely exchanged their
$\operatorname*{TE}\nolimits_{1}$ modes by using $N-1$ DCs. We obtain the
output fields,
\begin{equation}
\left\vert \psi_{1}^{\prime}\right\rangle =e^{i\phi_{1}}\binom{C_{0}^{1}%
}{C_{1}^{2}e^{i\lambda_{1}}},\left\vert \psi_{2}^{\prime}\right\rangle
=e^{i\phi_{2}}\binom{C_{0}^{2}}{C_{1}^{3}e^{i\lambda_{2}}},...,\left\vert
\psi_{N}^{\prime}\right\rangle =e^{i\phi_{N}}\binom{C_{0}^{N}}{C_{1}%
^{1}e^{i\lambda_{N}}}, \label{e13}%
\end{equation}
where $\lambda_{i}=\phi_{i+1}-\phi_{i},\lambda_{N}=\phi_{1}-\phi_{N},\left(
i=1...N-1\right)  $ are the phase differences, and $C_{0,1}^{i},\phi_{i}$ are
the mode coefficients and the random phases of the optical fields,
respectively. Then the output fields are employed to measure a small phase
difference $\theta$, and the output states can be written as,
\begin{equation}
\left\vert \psi_{1}^{\prime}\right\rangle =e^{i\phi_{1}}\binom{C_{0}^{1}%
}{C_{1}^{2}e^{i\left(  \lambda_{1}+\theta\right)  }},\left\vert \psi
_{2}^{\prime}\right\rangle =e^{i\phi_{2}}\binom{C_{0}^{2}}{C_{1}%
^{3}e^{i\left(  \lambda_{2}+\theta\right)  }},...,\left\vert \psi_{N}^{\prime
}\right\rangle =e^{i\phi_{N}}\binom{C_{0}^{N}}{C_{1}^{1}e^{i\left(
\lambda_{N}+\theta\right)  }}. \label{14}%
\end{equation}
At last the fields are sent into the Y splitters, and the intensity
differences are measured by photoelectric detectors. The detected
photocurrents are passively subtracted and performed correlation analysis. The
correlation function of the intensity differences can be obtained
\begin{align}
S\left(  \theta\right)   &  =\left\langle \hat{A}_{1}\left(  \theta\right)
\hat{A}_{2}\left(  \theta\right)  ...\hat{A}_{N}\left(  \theta\right)
\right\rangle \label{e15}\\
&  =\int_{\Lambda}%
{\displaystyle\prod\limits_{i=1}^{N}}
A_{i}\left(  \theta,\lambda_{i}\right)  \Phi_{i}\left(  \lambda_{i}\right)
d\lambda_{i}\nonumber\\
&  =\int_{\Lambda}%
{\displaystyle\prod\limits_{i=1}^{N}}
\cos\left(  \theta+\lambda_{i}\right)  \Phi_{i}\left(  \lambda_{i}\right)
d\lambda_{i}\nonumber\\
&  =\frac{1}{2^{N-1}}\cos\left(  N\theta\right)  ,\nonumber
\end{align}
where $\Phi_{i}\left(  \lambda_{i}\right)  $ are the distribution functions of
$\lambda_{i}$. The normalization factor $1/2^{N-1}$ can be removed by proper
normalization procedure. As before, the correlation function $S\left(
\theta\right)  $ can be estimated with an error $\Delta^{2}S\left(
\theta\right)  =\left[  \cos\left(  2N\theta\right)  -\sin^{2}\left(
N\theta\right)  \right]  /2^{2N-2}$. This means that the phase $\theta$ will
have an error $\Delta\theta=\Delta S\left(  \theta\right)  /\left\vert
\frac{\partial S\left(  \theta\right)  }{\partial\theta}\right\vert =1/N$,
that is the Heisenberg limit. This is a $\sqrt{N}$ enhancement over the
precision of $N$ measurements on independent fields.

In order to avoid the influence of the independent photon number distributions
of the independent fields, we can split one classical field into multiple
beams, then modulate each beam by an independent random phase shift. And the
multiple beams can be employed as multiple independent fields. Moreover the
granularity of random phases hardly influences the measurement result
\cite{Fu1}, so that the random phases can be assigned with a finite number of
discrete values uniformly distributed in $\left[  0,2\pi\right]  $ to realize
rapid phase ergodicity. In the scheme, the phase ergodicity might be one of
the most important sources of imprecision.

In this letter, we have discussed a novel interferometer by using optical
transverse modes in multimode waveguide that can beat the standard quantum
limit. By using a new linear optical scheme, $N$-field mode-entangled states
can be generated. Similar to the quantum-enhanced measurements, the $N$-field
mode-entangled states have achieved $\sqrt{N}$ enhancement over the precision
of the measurement $N$ times for independent fields. Compare to the quantum
scheme, the scheme can be realized much more easily. Although, we have
employed the interferometer using optical transverse modes, the results are
generally applicable to other types of interferometers, in particular to the
arrangements under development for gravity-wave detection.

This work was supported by the National Natural Science Foundation of China
under Grant No. 60407003.

\begin{description}
\item \pagebreak

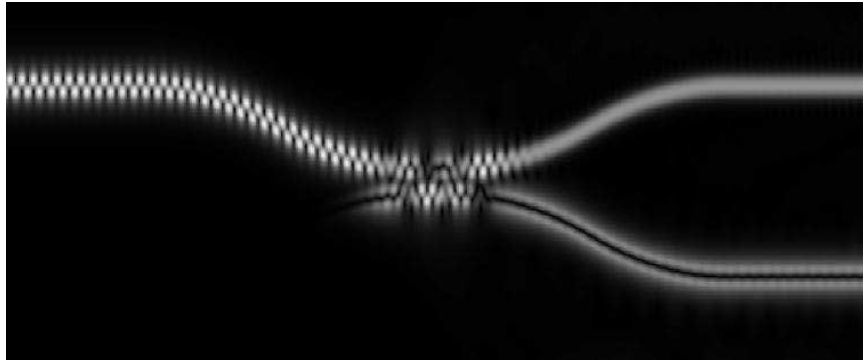
\item[Fig. 1:] BPM simulation result for the directional coupler to realize
mode separating.

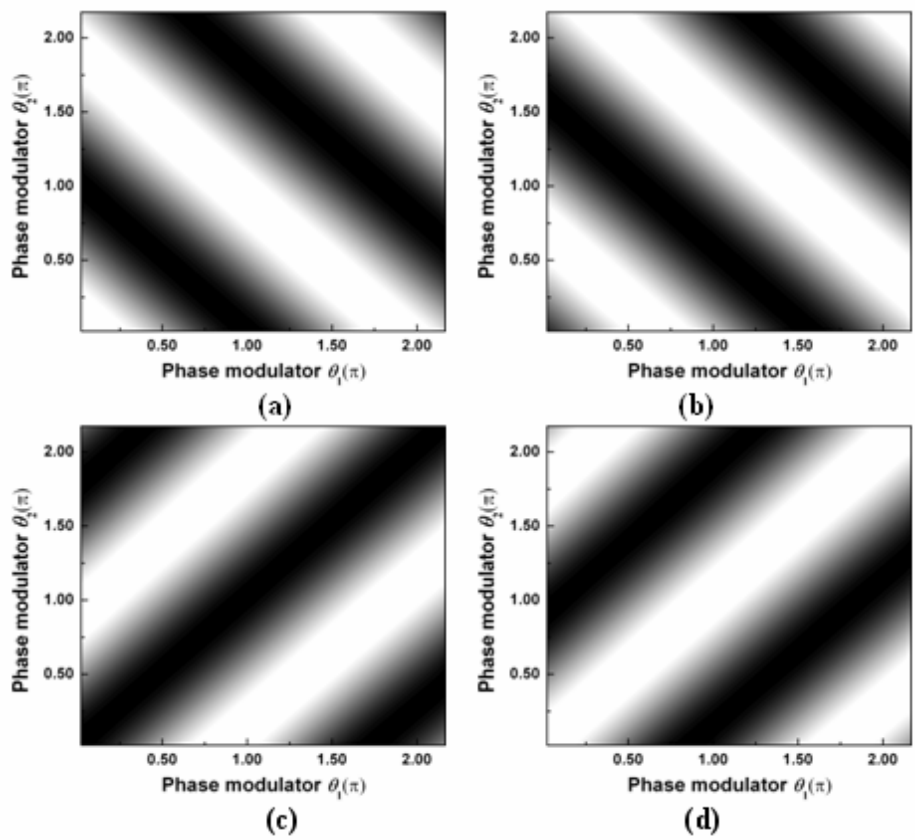
\item[Fig. 2:] The correlation functions $S\left(  \theta_{1},\theta
_{2}\right)  $ for mode-entangled states: (a) $\left\vert \Phi_{1}%
^{+}\right\rangle $, (b) $\left\vert \Phi_{1}^{-}\right\rangle $, (c)
$\left\vert \Psi_{1}^{+}\right\rangle $, and (d) $\left\vert \Psi_{1}%
^{-}\right\rangle $.

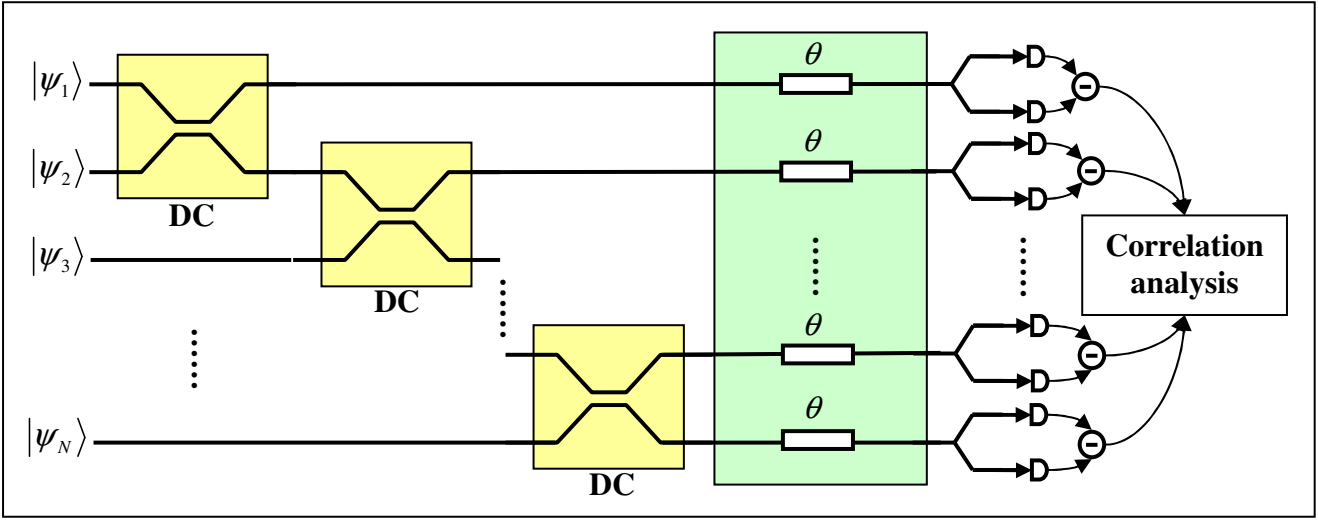
\item[Fig. 3:] The scheme of the interferometer.

\item[Tabel 1:] 

\item
\begin{tabular}
[c]{|l|l|l|}\hline
$\left\vert \Phi_{1}^{+}\right\rangle $ & $\theta_{1}=\frac{12}{46}\pi
,\theta_{1}^{\prime}=\frac{73}{46}\pi,\theta_{2}=\frac{66}{46}\pi,\theta
_{2}^{\prime}=\frac{5}{46}\pi$ & max$\left\vert B\right\vert =2.8174$\\\hline
$\left\vert \Phi_{1}^{-}\right\rangle $ & $\theta_{1}=\frac{39}{46}\pi
,\theta_{1}^{\prime}=\frac{19}{46}\pi,\theta_{2}=\frac{24}{46}\pi,\theta
_{2}^{\prime}=\frac{44}{46}\pi$ & max$\left\vert B\right\vert =2.8222$\\\hline
$\left\vert \Psi_{1}^{+}\right\rangle $ & $\theta_{1}=\frac{39}{46}\pi
,\theta_{1}^{\prime}=\frac{100}{46}\pi,\theta_{2}=\frac{88}{46}\pi,\theta
_{2}^{\prime}=\frac{68}{46}\pi$ & max$\left\vert B\right\vert =2.8152$\\\hline
$\left\vert \Psi_{1}^{-}\right\rangle $ & $\theta_{1}=\frac{38}{46}\pi
,\theta_{1}^{\prime}=\frac{18}{46}\pi,\theta_{2}=\frac{87}{46}\pi,\theta
_{2}^{\prime}=\frac{67}{46}\pi$ & max$\left\vert B\right\vert =2.8218$\\\hline
\end{tabular}

\end{description}

\end{document}